 \documentclass[iop]{emulateapj}
 \pdfoutput=1

\usepackage{color}
\usepackage{amsbsy}
\usepackage{amssymb}
\usepackage{amsmath}
\usepackage{bm}
\usepackage{graphicx}

\def\vec#1{\ensuremath{\mathchoice{\mbox{\boldmath$\displaystyle#1$}}
{\mbox{\boldmath$\textstyle#1$}}
{\mbox{\boldmath$\scriptstyle#1$}}
{\mbox{\boldmath$\scriptscriptstyle#1$}}}}

\slugcomment{ Accepted to ApJ: October 6, 2014}

\shorttitle{Oscillations and Convection}
\shortauthors{Antoci et al.}

\begin{document}
\bibpunct{(}{)}{;}{a}{}{,}
%% LaTeX will automatically break titles if they run longer than
%% one line. However, you may use \\ to force a line break if
%% you desire.

\title{The role of turbulent pressure as a coherent pulsational driving mechanism: \\
 the case of the $\delta$\,Scuti star HD 187547 }

%% Use \author, \affil, and the \and command to format
%% author and affiliation information.
%% Note that \email has replaced the old \authoremail command
%% from AASTeX v4.0. You can use \email to mark an email address
%% anywhere in the paper, not just in the front matter.
%% As in the title, use \\ to force line breaks.

\author{V. Antoci\altaffilmark{1}, M. Cunha\altaffilmark{2}, G. Houdek\altaffilmark{1}, H. Kjeldsen\altaffilmark{1}, R. Trampedach,\altaffilmark{1,3}, G. Handler\altaffilmark{4}, \\
T. L\"uftinger\altaffilmark{5}, T. Arentoft\altaffilmark{1}, S. Murphy\altaffilmark{1, 6} }
\affil{\altaffilmark{1} Stellar Astrophysics Centre,  Aarhus University, Ny Munkegade 120, DK-8000 Aarhus C, Denmark; antoci@phys.au.dk}
\affil{\altaffilmark{2}Centro de Astrof\'{i}sca e Faculdade de Ci\^{e}ncias, Universidade do Porto, Rua das Estrelas 4150-762, Portugal}
\affil{\altaffilmark{3}JILA, University of Colorado and National Institute of Standards and Technology, 440 UCB, Boulder, CO 80309, USA}
\affil{\altaffilmark{4}Copernicus Astronomical Center, Bartycka 18, 00-716 Warsaw, Poland}
\affil{\altaffilmark{5}Institute for Astronomy, University of Vienna, T\"urkenschanzstr. 17, A-1180 Vienna, Austria}
\affil{\altaffilmark{6}Sydney Institute for Astronomy (SIfA), School of Physics, University of Sydney, 2006 Australia}

\begin{abstract}
HD\,187547 was the first candidate that led to the suggestion that solar-like oscillations are present in $\delta$\,Scuti stars. Longer observations, however, show that the modes interpreted as solar-like oscillations have either very long mode lifetimes, longer than 960 days, or are coherent. These results are incompatible with the nature of `pure' stochastic excitation as observed in solar-like stars. Nonetheless, one point is certain: the opacity mechanism alone cannot explain the oscillation spectrum of HD\,187547. Here we present new theoretical investigations showing that convection dynamics can intrinsically excite coherent pulsations in the chemically peculiar $\delta$\,Scuti star HD\,187547. More precisely, it is the perturbations of the mean Reynold stresses (turbulent pressure) that drives the pulsations and the excitation takes place predominantly in the hydrogen ionization zone. 
\end{abstract}


\section{Introduction}

Stars of spectral type A and early F cover a very interesting part of the Hertzsprung Russell Diagram (HRD) where several astrophysical processes interact. The classical instability strip crosses the granulation boundary between the deep envelopes of efficient convection on the cool side, to shallow, inefficient convective layers barely able to transport flux, on the warm side. How this transition takes place is not entirely understood. It is of great interest, however,  to infer how the stellar structure changes with increasing mass, because convection in the outer layers affects not only the stellar structure, but also the dynamo-generated magnetic fields, activity and angular momentum transport, etc.

The lower part of the classical instability strip is populated by several classes of pulsators: $\delta$\,Scuti ($\delta$\,Sct) stars, $\gamma$ Doradus ($\gamma$\,Dor), rapidly oscillating Ap (roAp) stars, but also by non-pulsating stars. The main mechanism which is commonly accepted to be responsible for the excitation of pulsations in $\delta$\,Sct stars is the $\kappa$ (opacity) mechanism acting like a heat engine in the He\,II ionization zone, which is located at a layer with temperature  $T \backsimeq$ 40\,000\,K. However there is also a small but measurable contribution from the thin, combined hydrogen (H\,I) and first ionization zone of helium (He\,I) to the excitation of oscillations (Castor 1968). This contribution results in an observed phase lag between maximum light and the time of minimum radius obtained from radial velocity measurements. In $\delta$\,Sct stars  this phase lag is of the order of 60$\degr$, considerably smaller than the 90$\degr$ in classical Cepheids (Breger, Hutchins \& Kuhi, 1976). This is because the H\,I and He\,I ionization zones are less efficient at driving low-order modes and are closer to the surface in $\delta$\,Sct stars compared to classical Cepheids. 

Houdek et al. (1999) and Samadi et al. (2002) predicted that the subsurface convection in stars in the classical instability strip is still vigorous enough to stochastically excite solar-like oscillations. The characteristics of solar-like oscillations permit identification of the geometry of oscillation modes from pattern recognition methods, e.g., from echelle diagrams. This is possible because all modes in a certain frequency range are excited and visible. This is in absolute contrast to oscillations excited by any heat engine mechanism, such as the $\kappa$ mechanism, where the mode selection mechanism is not yet understood (see, e.g., the review by Smolec 2014). The  same  applies to the oscillation amplitudes. While for solar-like oscillations we can predict their approximate value (e.g. Houdek et al. 1999), this is not the case for heat-engine driven pulsators, where non-linear non-adiabatic codes are needed to estimate their amplitudes.\par

\begin{figure*}[t!]
\begin{center}
\includegraphics[width=7.2cm]{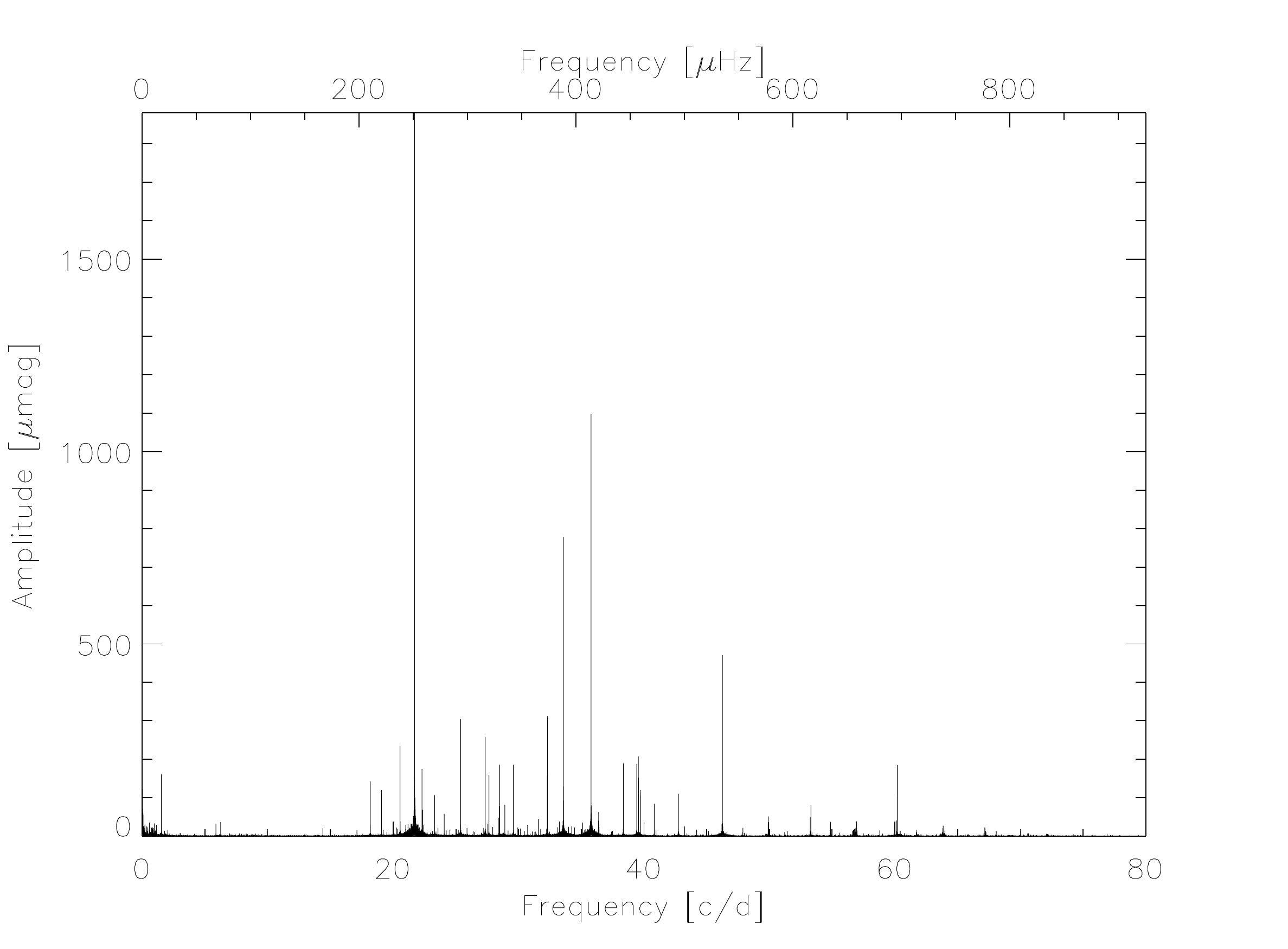}
\includegraphics[width=7.2cm]{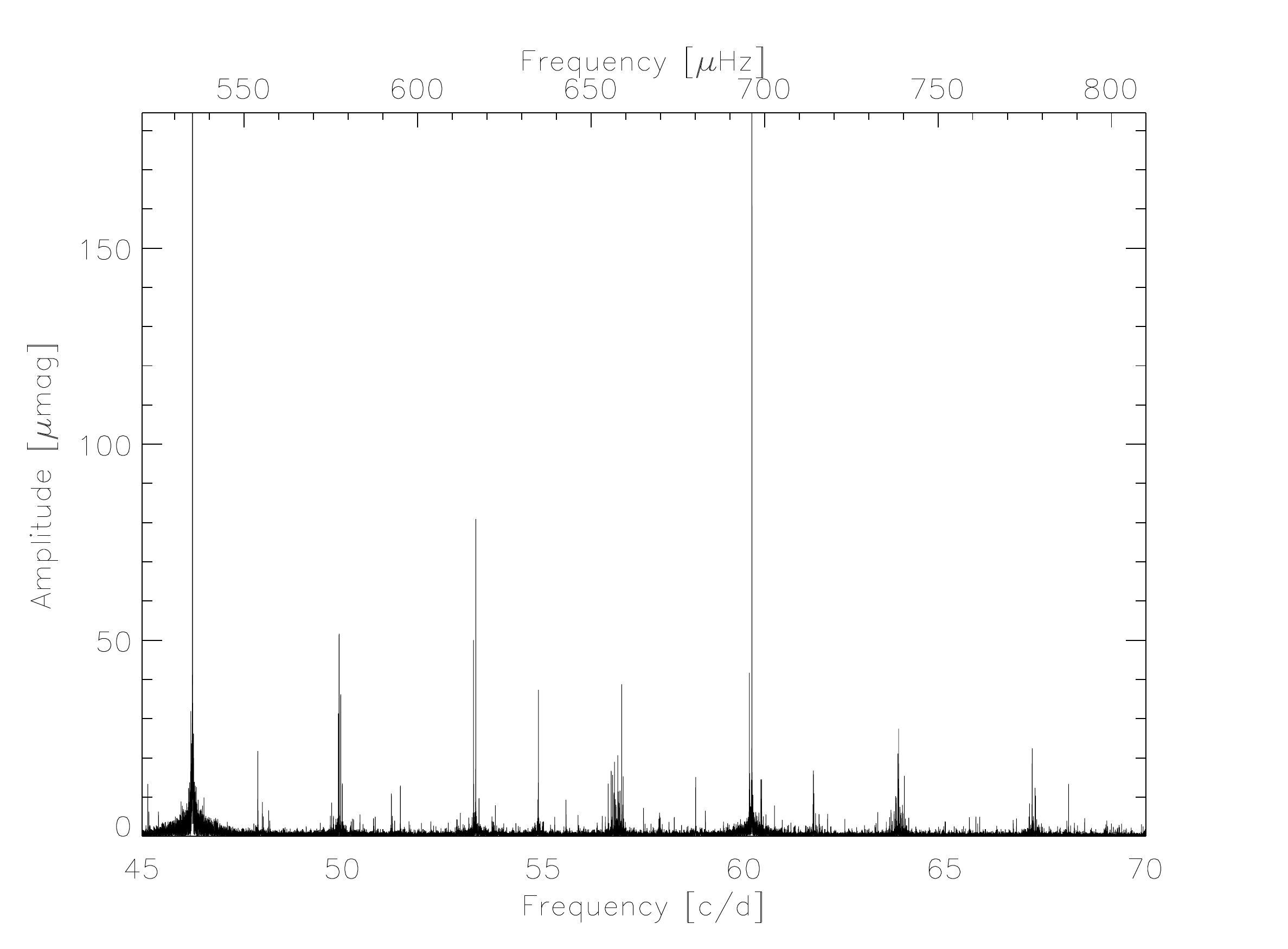}
\caption{Left panel: Fourier spectra of the Kepler short cadence data. Right panel: Close-up in the frequency region interpreted by Antoci et al. (2011) to be stochastically excited. \label{fou} }
\end{center}
\end{figure*}

HD\,187547, a $\delta$\,Sct star detected with the NASA spacecraft {\it Kepler}, was the very first of its kind to lead to the conclusion that solar-like oscillations were indeed present in $\delta$\,Sct stars  (Antoci et al. 2011), as predicted by theory (Houdek et al. 1999, Samadi et al. 2002). The unusual oscillation spectrum of this star (Fig.~\ref{fou}) shows a very large range of excited pressure modes at low and high radial orders.  Antoci et al's (2011)  interpretation was based on the striking similarities between the observations and the characteristics of stochastic oscillations: (1) The frequency range observed in HD\,187547 cannot be explained by the opacity mechanism as we understand it. (2) The observed pattern at the high-frequency modes is consistent with the large frequency separation, $\Delta \nu$, expected for high-radial orders. (3)  The scaling relation between the large frequency separation, $\Delta \nu$, and the frequency of maximum power, $\nu_{\rm max}$, correctly predicts the mode with the highest amplitude within the range of modes interpreted as solar-like oscillations (e.g., Huber et al. 2011). (4) The broad peaks are observed only around the modes of high radial orders (Solar-like oscillations are non-coherent because they are intrinsically damped and re-excited oscillations. Opacity driven pulsations, on the other hand seem to be stable and coherent over very long time scales. Coherent and non-coherent signals can be distinguished in Fourier spectra by comparing  the width and the shape of the peak to the window function). (5) The statistical properties of the high-order modes were comparable with those of stochastically excited oscillations (Chang \& Gough 1998). This was not the case for the low-frequency modes, excited by the $\kappa$ mechanism. \par
Abundance analyses of HD\,187547 revealed that the star is a chemically peculiar Am star, displaying photospheric overabundances in Ba, Y and Sr and underabundances in Sc and Ca (Preston 1974).  According to Auriere et al. (2011) these stars do not have any large-scale magnetic fields, also confirmed in the case of HD\,187547, contrary to the likewise chemically peculiar (ro)Ap stars. The Am phenomenon is related to atomic diffusion which can efficiently operate because of slow rotation (Charbonneau 1993). The pulsating AmFm stars still represent a mysterious group, as the reason why a large fraction pulsates is not well understood. It is believed that, due to settling of He, this element is not sufficiently abundant in the He\,II zone for the $\kappa$ mechanism to drive pulsation. The implementation of diffusion of heavy elements in models can explain pulsation in a very constrained instability region (Turcotte et al. 2000). However, observations from the {\it Kepler} spacecraft and from ground (Balona et al. 2011, Smalley et al. 2013) demonstrate that pulsating AmFm stars are spread over the entire $\delta$\,Sct instability strip. In other words HD\,187547 not only has an unusually large range of pulsation modes excited for a `normal' $\delta$\,Sct star but  even more so for a chemically peculiar Am star.\par

\begin{figure*}[htb!]
\begin{center}

\includegraphics[width=12cm, trim=0cm 0.5cm 0cm 0cm]{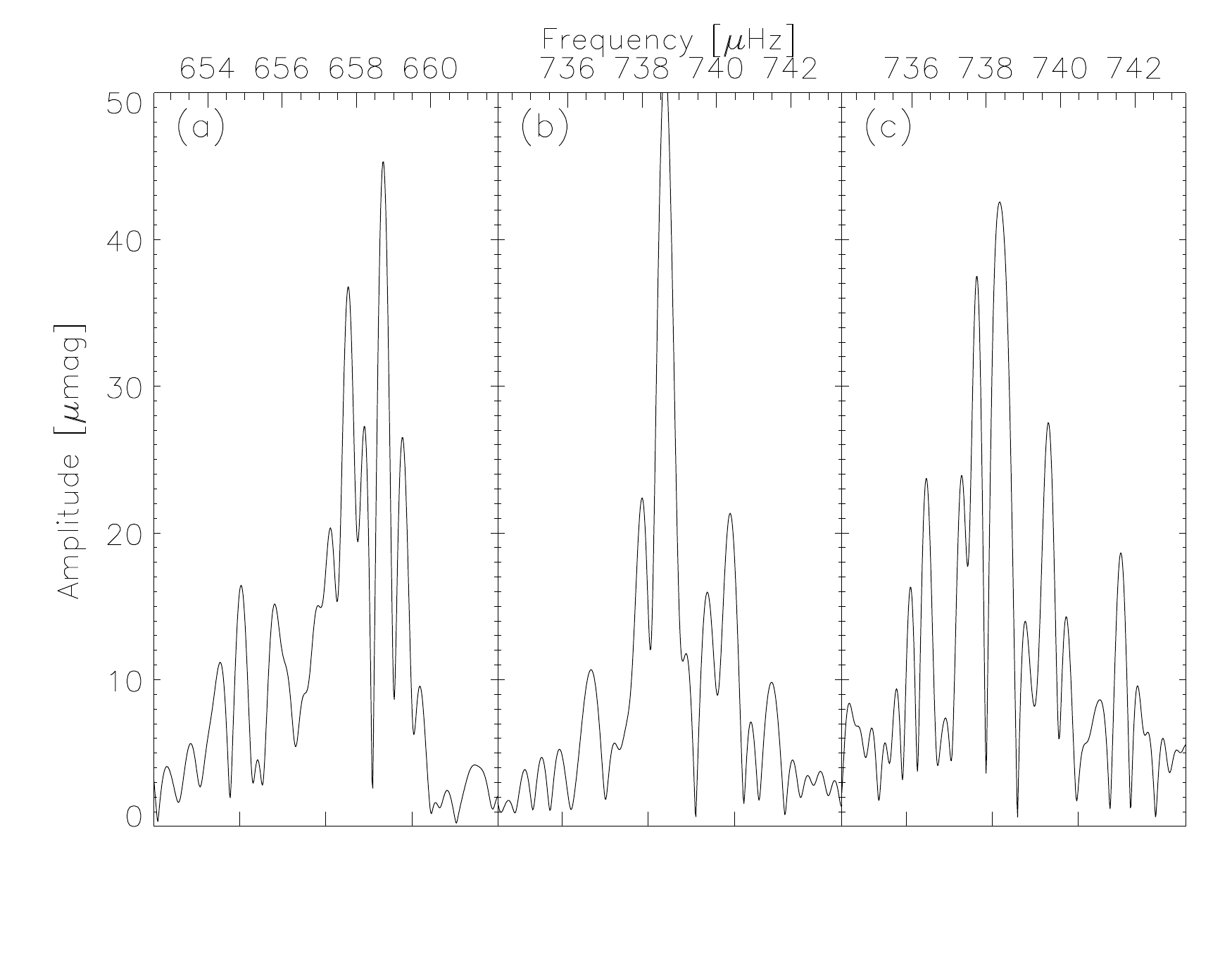}

\includegraphics[width=12cm, trim=0cm 0cm 0cm 1.5cm]{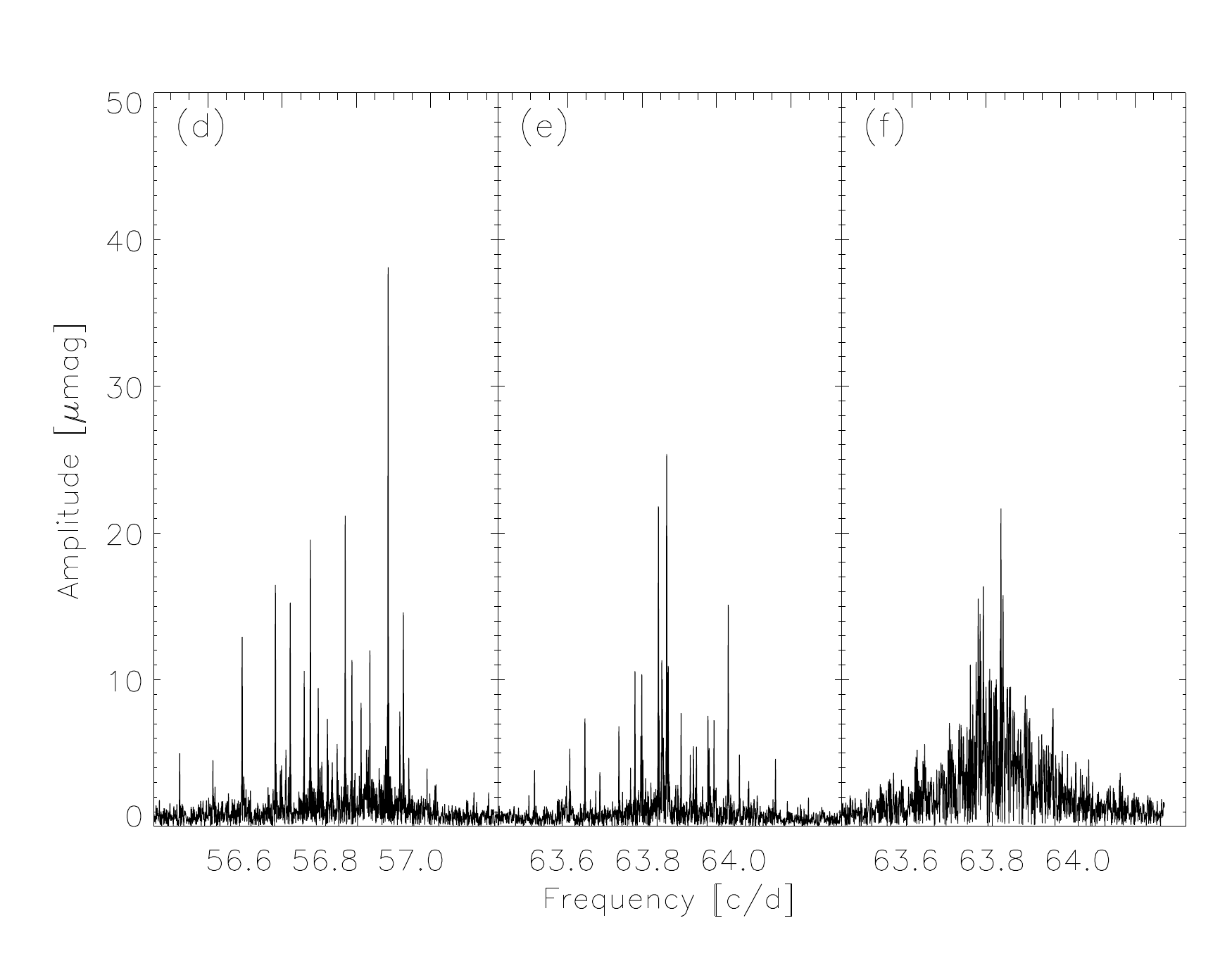}

\caption{In panels (a), (b), (d) and (e)  we show Fourier spectra of two oscillation modes observed in HD\,187547 and interpreted by Antoci et al (2011) as being stochastically excited. Panels (c) and (f) depict a simulated stochastic,  damped and re-excited,  oscillation mode. The Fourier spectra in panels (a), (b), (c) are based on 30 days of observations (quarter 3.2), the same data set  used by Antoci et al (2011). Panels (d) and (e) illustrate the oscillation spectra of HD\,187547 based on 960 days of short cadence data, clearly showing the temporally stable and well resolved peaks.  For comparison, in panel (f) we show the Fourier spectrum of a simulated stochastically excited, i.e. non-coherent mode with a mode lifetime of 2.8 days using the same duration for the time series as in panels (d) and (e). It can be seen that for a stochastic signal the amplitude decreases with increasing observing time, a characteristic we do not see for the coherent modes, unless beating due to unresolved peaks occurs as illustrated for the mode in panel (b) and (e).  \label{mode}}
\end{center}

\end{figure*}

\section{Data analyses and interpretation}

The {\it Kepler} spacecraft launched in 2009 (Koch et al. 2010) observed our target HD\,187547 (KIC\,7548479, RA$_{2000}$=19\,48\,36.5, DEC$_{2000}$=43\,06 32.3) for 960 days at short-cadence (1-min cadence for  quarters 3.2 and 7 to 17) and 1470 days at long-cadence observing mode (30-min cadence for quarters 0-17).\par

 In Fig.~\ref{fou} we show a Fourier spectrum of the entire data set, together with a close-up of the frequency range initially interpreted to be stochastically excited. With more than two years of data it is now obvious that what Antoci et al. (2011) suggested to be broad peaks with short mode lifetimes based on one month of data, as expected for solar-like oscillations, are in fact very sharp, well separated and individually coherent, but closely spaced peaks as shown in Fig.~\ref{mode}. The upper row of Fig.~\ref{mode} shows the Fourier spectra of three modes computed from 30 days of data, two of which were observed in HD\,187547 (panels (a) and (b)) and a simulated stochastic mode (panel (c)), demonstrating the similar structure and appearance between the observed and the simulated modes (see also fig.\,2 from Antoci et al. 2011). The lower row of Fig.~\ref{mode} illustrates the same modes, but this time computed from 960 d of data. It is clear that the peaks shown in panels (d) and (e) are stable and well resolved, which is in absolute contrast to the stochastic mode in panel {\bf (f)}. The latter shows unresolved peaks and a significantly lower peak height typical for a signal being damped and re-excited.

 {\it The stable temporal behaviour of the observed modes is not consistent with stochastic excitation}, unless the mode lifetimes are longer than the observations, for which the widths of the peaks and the temporal stability provide no evidence as seen in  Fig.~\ref{mode}. Furthermore, observations of solar-like oscillators clearly show that the mode lifetimes decrease with increasing effective temperatures, and is of the order of one day for the hottest stars (e.g. White et al. 2012 and references therein). 
 While the observed modes exclude stochastic excitation as seen in sun-like stars, these do not  explain the mechanism triggering the oscillations. \par

A closer look at the Fourier spectrum reveals an agglomeration of peaks around several of the modes at high radial orders, which is expected for modes of the same degree $l$ of consecutive radial orders. The observed frequency separation (3.5cd$^{-1}$; $40.5\mu$Hz) is consistent with the large frequency separation $\Delta \nu$ for a star with the parameters of HD\,187547, i.e. $T_{\rm eff}$=7500$\pm$200K and $\log g$=3.9$\pm$0.2 dex. However, there are some problems explaining the peaks around some of the modes (e.g. Fig.~\ref{mode}, panels (d) and (e)), which we will address in the next paragraphs.

%\newpage

{\bf Rotational splitting of modes}: 
Rotational splitting {\it alone} cannot explain the above mentioned agglomeration of peaks. Even under the assumption that {\it all} modes with $l\le3$ as well as their associated rotational splittings have observable amplitudes, which is actually unlikely due to geometrical cancellation effects (e.g. Aerts et al. 2010), there are still too many peaks to be accounted for around some of the modes (see Fig.\ref{mode}).  Additionally, several of the splittings are too small to be caused by rotation, given the measured $v \sin i$ of 11 $\pm$ 1 kms$^{-1}$ (Antoci et al. 2011). The Am-nature of HD\,187547 and the low projected rotational velocity suggest that the star is a slow to moderate rotator, for which fairly equidistant patterns for rotationally split modes associated with modes of degrees 1, 2 and 3 are expected. We used the autocorrelation technique to search for patterns in the smoothed Fourier spectrum of HD\,187547. We subdivided the Fourier spectrum into a low (20 -- 45cd$^{-1}$) and a high (45 -- 70cd$^{-1}$) frequency domain, and checked the frequency region around each high-radial order mode separately. No clear dominating or re-occurring pattern, i.e. no re-occurring equidistant peaks,  can be found, neither around the separate modes nor in the low or high-frequency domain. Hence we conclude that rotational splitting {\bf alone} is not a viable explanation for the agglomeration, but we do expect that some peaks are indeed rotationally split modes.

{\bf High-degree modes:}
Given the high quality of the {\it Kepler} data it is plausible that modes of higher degrees ($l\ge4$) are also observed, even if their amplitudes are heavily suppressed due to cancellation effects. Following Ballot et al. (2011) and Lund et al. (2014) we calculate the visibilities with respect to $l=0$ for modes with degrees from $l=0$ to $l=6$ to be 1, 1.48,   0.49,  0.018,  0.0073, 0.0011, 0.00095 respectively. Assuming that one of the modes with the highest amplitudes is a radial or dipolar mode and adopting the same intrinsic amplitude for all modes, we find it unlikely but cannot rule out that we observe $l>4$ modes. While $l=4$ modes might explain the number of observed peaks, these do not explain the observed agglomeration of peaks, as $l=4$ modes are not expected to align with any of the modes of degree $l\le3$ in the asymptotic regime, i.e. at high radial orders. The $l=4$ modes are found in between the ridges of the $l=0,2$ and $l=1,3$ modes, not only for a model with the parameters of HD\,187547 but also for the Sun (Lund et al. 2014). \par

{\bf Mixed modes:}
Mixed modes, i.e. modes behaving as gravity modes in the stellar interior and as pressure modes in the outer envelope, are expected to be observed in pulsating stars as they evolve. However, the evolutionary stage of HD\,187547 is not advanced enough (see Antoci et al. 2011) to explain the large number of peaks observed in HD\,187547 by mixed modes as found in red giants (e.g. Bedding et al. 2011)

{\bf Strong magnetic fields:} Strong magnetic fields, as observed in roAp stars could cause rotational amplitude modulation of modes that are aligned with the magnetic axes and not with the rotational axes (oblique pulsator model, Kurtz 1982). This scenario will result in additional peaks in a Fourier spectrum. To investigate, whether HD\,187547
possesses a magnetic field strong enough to cause an observable mode splitting, we obtained spectropolarimetric measurements of the star on 2012 October 3 using the NARVAL echelle spectrograph in polarimetric mode during DDT time of the T\'elescope Bernard
Lyot (Observatoire du Pic du Midi, France). The spectropolarimeter offers full optical wavelength coverage from 3700 - 10\,000\AA~  in a single exposure with a resolving power of approximately 65\,000. The data were reduced using the LIBRE-ESPRIT automatic reduction software package for spectropolarimetric data, which is based on the ESPRIT reduction package (described in detail in Donati et al. 1997).
As the S/N is too low to detect Zeeman signatures in individual lines, we applied the Least-Squares Deconvolution (LSD) technique,
(Donati et al. 1997, Kochukhov 2010), but did not find any polarimetric signal originating from a magnetic field larger than 20G.
Such weak magnetic fields cannot split pulsation modes as observed in roAp stars. 
\par

{\bf Combination frequencies:} Many $\delta$\,Sct stars show combination frequencies of the form, $mf_i\pm nf_j$, where $m$ and $n$ are integer numbers and $f_{ij}$  the observed frequencies, in their oscillations spectra, as a result of the pulsational being  non-linear. 
In the case of HD\,187547 we find several hundreds statistically significant peaks, which makes it hard to distinguish clearly between real combination modes and accidental matches.  We checked two scenarios, in which we required the parent modes to have an amplitude higher than 50 $\mu$mag and 10 $\mu$mag, respectively.  We conclude that most of the peaks at high frequencies cannot be reproduced by second order combination frequencies (i.e.  $f_i+f_j$), which are the most likely to occur (Papics 2012). Nevertheless as a test, we pre-whitened peaks with the same values as second order combination frequencies, and again autocorrelated the Fourier spectrum and found no clear pattern due to rotationally split modes.     \par

{\bf Companion:} If the star were in a binary or multiple system, it would have a variable velocity with respect to Earth. This would cause a Doppler shift of each and every pulsation mode, which would be observed as a frequency splitting equal in magnitude to the orbital frequency of the system (Shibahashi \& Kurtz 2012). Such orbital frequency splittings are not observed for HD\,187547. Additionally we searched for pulsational phase modulation, following the method of Murphy et al. (2014). We used the five highest peaks in the Fourier spectrum, namely those at 21.7, 35.8, 33.6, 25.4 and 20.6\,cd$^{-1}$, and determined their frequencies, amplitudes and phases with a non-linear least-squares fit to the entire Kepler long cadence data set (quarter 0 to 17). We then subdivided the time-series into 10-d segments, and recalculated the phase at fixed frequencies in each segment. Let us assume the pulsating star is in a binary system. Then we may attribute periodic phase variations to variations in the light arrival time because the path length travelled by the light on its way to Earth varies. We see no periodic variations in phase and thus no periodic variation in the light arrival time delays. None the less, from a Fourier transform of the light arrival time delays, we are able to place an upper limit on a hypothetical binary mass function, of $1.1\times10^{-5}$\,M$_{\odot}$. If we assume a primary mass of 1.85\,M$_{\odot}$ and an arbitrary inclination angle $i=45^{\circ}$, we are able to rule out a companion more massive than 0.049\,M$_{\odot}$. To provide an upper limit on companion mass, we take the same mass function and assume the primary mass is at the high-mass end of the distribution of masses of $\delta$\,Sct stars, at 2.1\,M$_{\odot}$, and we assume $i=15^{\circ}$. With these values we can rule out a companion having a mass $m\geq0.15$\,M$_{\odot}$.

{\bf To conclude}, based on the temporal stability and the narrow peaks defined by the window function, the high radial order modes are most likely not stochastically excited. We find an agglomeration of peaks around some of the high radial order modes (Fig.~\ref{mode}), but cannot offer a clear explanation to account for all the peaks observed at high frequencies. We can, however, exclude strong magnetic fields to cause additional splitting of modes as observed in roAp stars as well as a companion with a mass larger than than 0.049\,M$_{\odot}$ to induce additional peaks due to the light time effect. We find very few peaks to have the same values as second order combination frequencies, which in case combinations frequencies are present are the most likely to occur. Based on the Am-nature of the HD \,187547 we can assume slow to moderate rotation resulting in more or less equidistantly spaces peaks, but we find no clear pattern to explain rotationally split modes, which does not mean that these are not present, only that they are not clearly identifiable. As far as high degree modes are concerned, we cannot exclude the presence of modes with $l\geq4$ but find them unlikely to explain the agglomeration round some of the high-order modes because of significant geometric cancellation.

\section{Excitation by the turbulent pressure mechanism}
 
In view of the latest Kepler data, which suggest that the high-frequency
modes observed in HD\,187547 are not stochastically excited, we perform a nonadiabatic 
pulsational stability analyses of two stellar models for this star with the same mass
and effective temperature, but different in luminosity. 
The global properties of the two models are listed in Table~1.
The linear stability analyses were carried out in the manner of
Balmforth (1992a) and Houdek et al. (1999) using, however, the latest OPAL
opacity tables (Roger \& Iglesias 1995) supplemented at low temperatures by 
the Ferguson et al. (2005) tables. In these analyses the turbulent fluxes are 
obtained from a nonlocal generalization (Gough 1977a) of Gough's (1977b) time-dependent 
mixing-length formulation, adopting for the mixing length a value that was calibrated
to the Sun. The momentum flux (turbulent pressure) is consistently implemented in both the equilibrium and linear stability calculations as described by Houdek et al. (1999). The stellar models described above are computed for radial modes only, however, we can assume that the results are also applicable for low-order non-radial modes as well. This is because most non-adiabatic processes take place in the outer thin layers of the star, where the structure of a low-degree modes differ very little from that of radial modes of similar frequencies  (as discussed by Balmforth et al. 2001). \par

In this nonlocal generalization there are three more parameters, $a$, $b$ and $c$, which control 
the spatial coherence of the ensemble of eddies contributing
to the turbulent fluxes of heat ($a$), and momentum ($c$), and the degree to which
the turbulent fluxes are coupled to the local stratification ($b$).
These parameters control the degree of `nonlocality'
of convection: low values imply highly nonlocal solutions, and in the limit
$a,b,c\rightarrow\infty$ the system of equations formally reduces to the local 
formulation (e.g., Houdek et al. 1999), except near the boundaries of 
the convection zone, where the local equations are singular. 
Theory suggests values for these nonlocal convection parameters (Gough 1977a), which
are of order unity, but these values are approximate and to some extent these
parameters are free. Balmforth (1992a) explored the effect of $a$, $b$ and $c$
on the turbulent fluxes for the solar case and found good agreement between measured
solar mode lifetimes and (twice the) estimated linear damping rates using  
$a^2=600$, $b^2=600$ and $c^2=600$.
Houdek \& Gough (2002) estimated linear damping rates in the red-giant star $\xi$\,Hydrae and
found mode lifetimes agreeing with observations for $a^2=900$, $b^2=2000$, and $c^2=300$.

\begin{figure}[t!]
\begin{center}
\includegraphics[width=8cm]{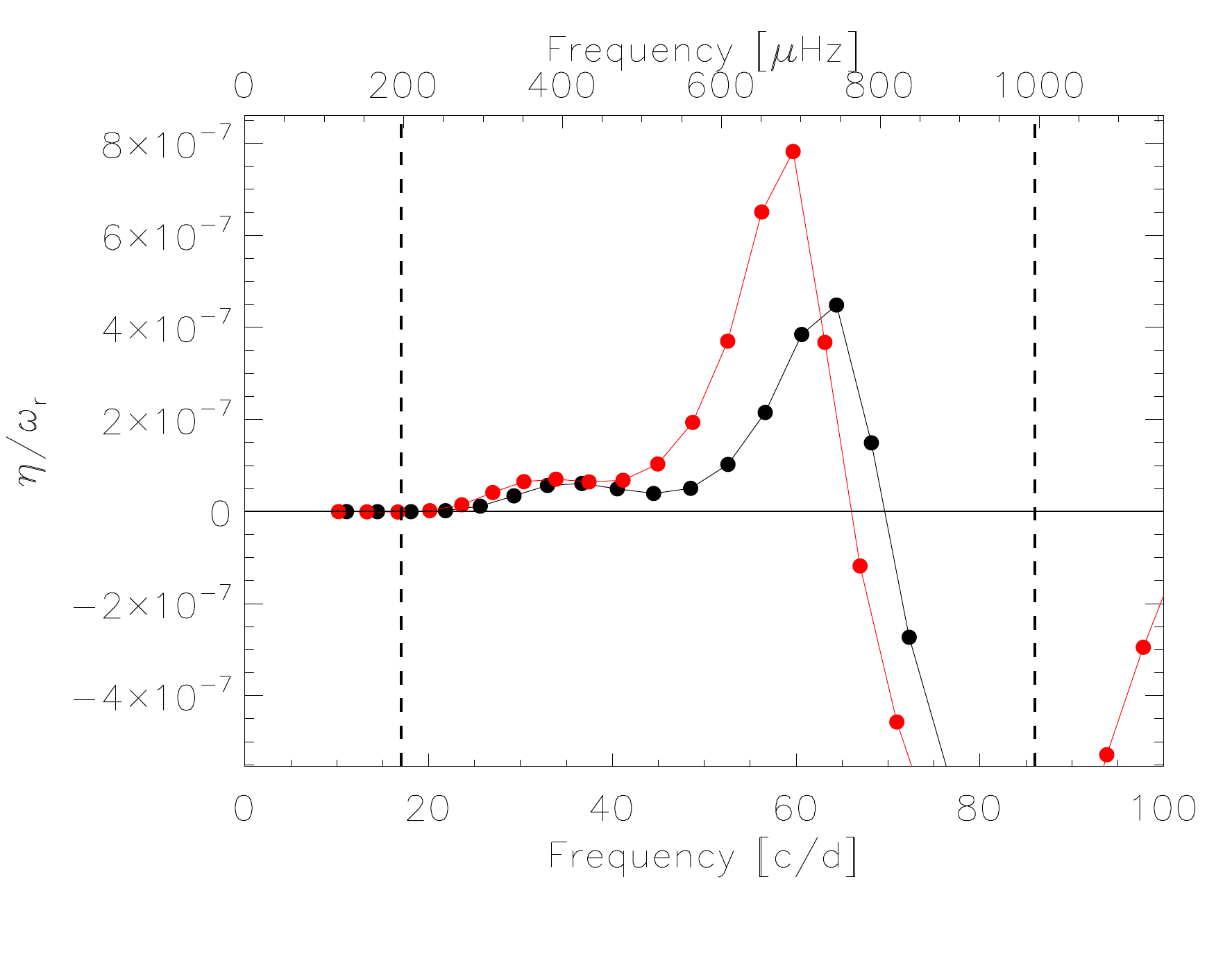}
\caption{Comparison between the theoretically excited range of modes (full circles) and the observed modes delimited by the dashed lines. Here we plot the normalized growth rate as a function of frequency for model 1 (black) and model 2  (gray/red), see Table~1 for details on parameters.  A positive growth rate means excited modes, whereas a negative value indicates intrinsically damped modes. \label{growth}}
\end{center}
\end{figure}

The same nonlocal convection model was also applied to roAp stars by
Balmforth et al. (2001), who adopted $a^2=1000$, $b^2=1000$ and $c^2=1000$
in the stability analyses of radial p modes. These rather large values for $a$, $b$ and $c$ indicate that convection is more local, which is not inconsistent with the shallow convective envelopes in A and early F-type stars. These values also reproduce observations of roAp stars best (see, e.g. Cunha 2013). In this paper we adopt values similar to Balmforth et al. (2001), i.e. $a^2=950$, $b^2=950$ and $c^2=950$, as such high values drive  a larger number of modes at higher frequencies, consistent with the observations presented here. Note that Antoci et al. (2011) adopted Balmforth's (1992a) solar-calibrated values for the their preliminary estimates of stability properties in HD\,187547 and did not find driving at high radial orders.

Our analyses are illustrated in Fig.~\ref{growth} for both models, where we plot the normalized growth rates, i.e., the linear stability coefficient, $\eta / \omega_{\rm r}$ as a function of frequency, where the complex angular frequency $\omega = \omega_{\rm r}+i \eta$. A positive growth rate, $\eta$, indicates intrinsically excited modes, whereas a negative growth rate indicates that the modes are intrinsically damped.  

\begin{figure*}[htb!]

\begin{center}

\includegraphics[width=8cm]{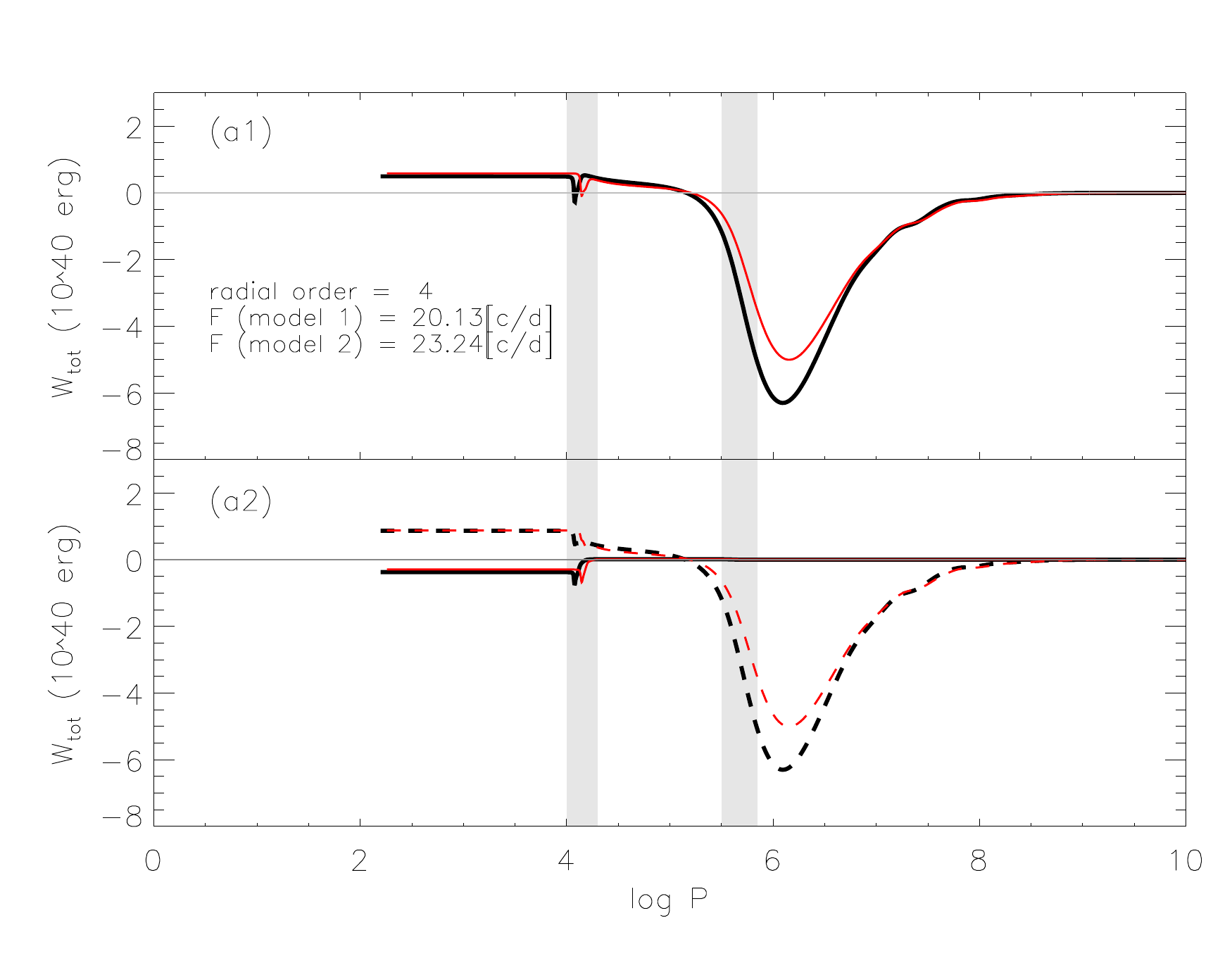}
\includegraphics[width=8cm]{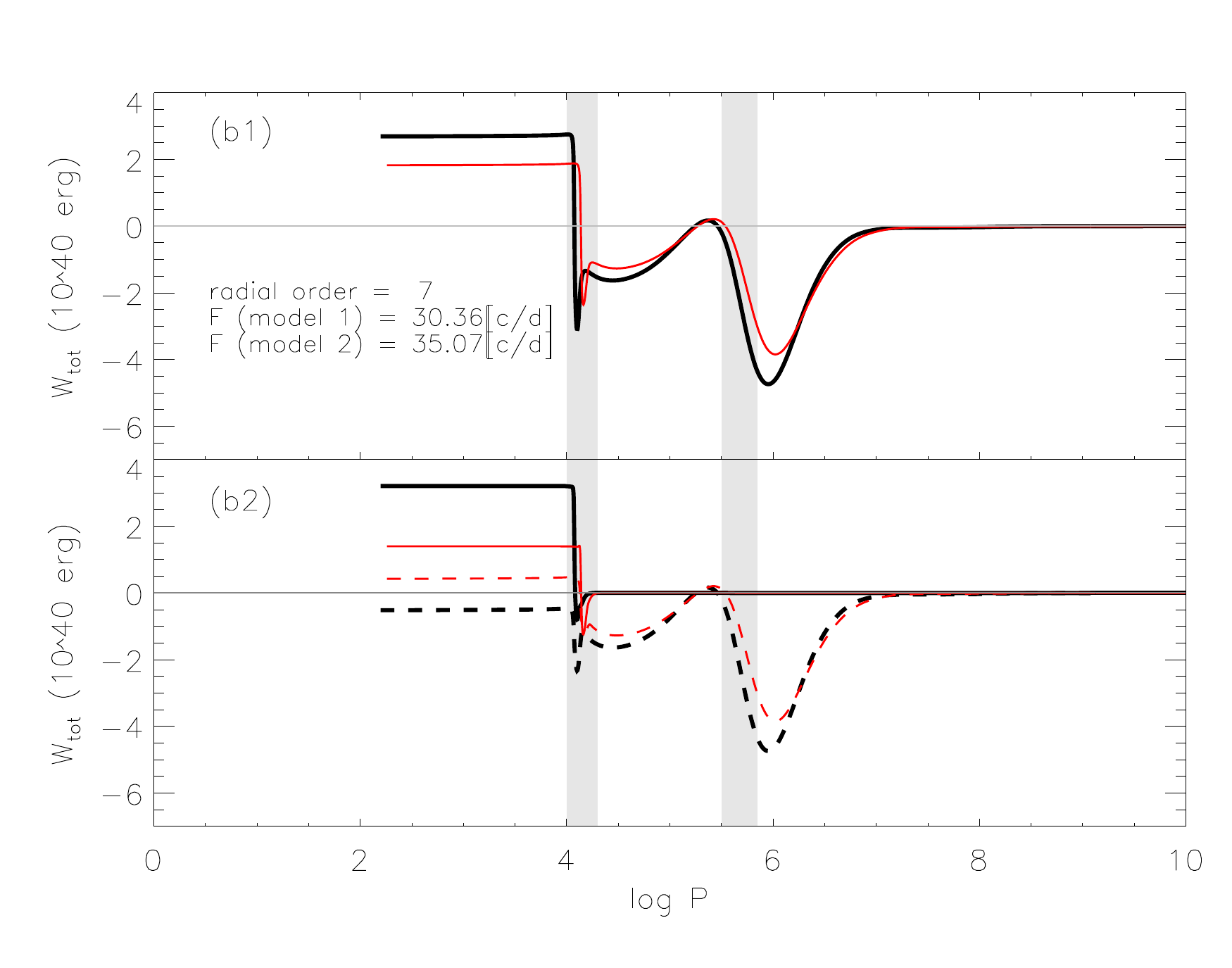}

\includegraphics[width=8cm]{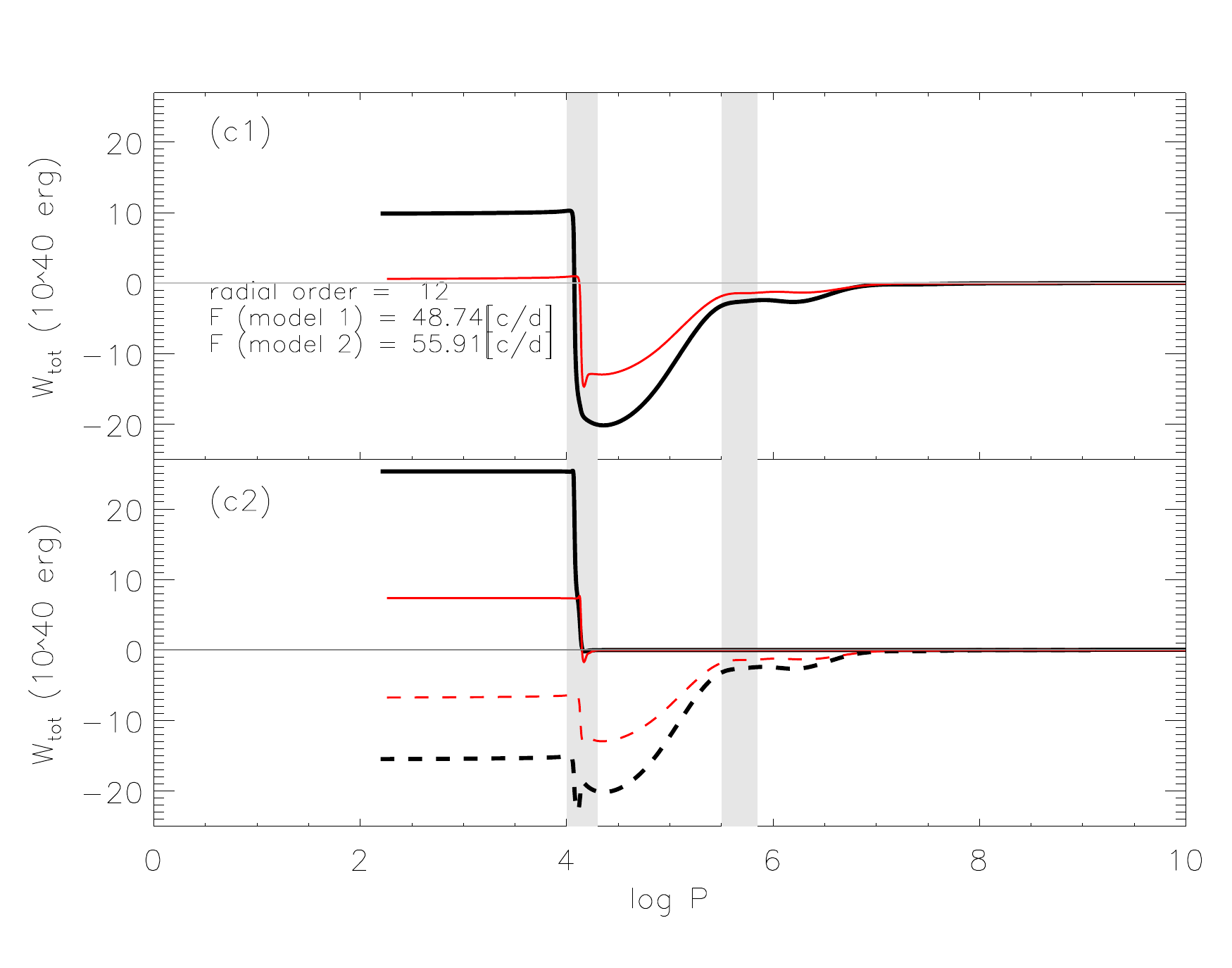}
\includegraphics[width=8cm]{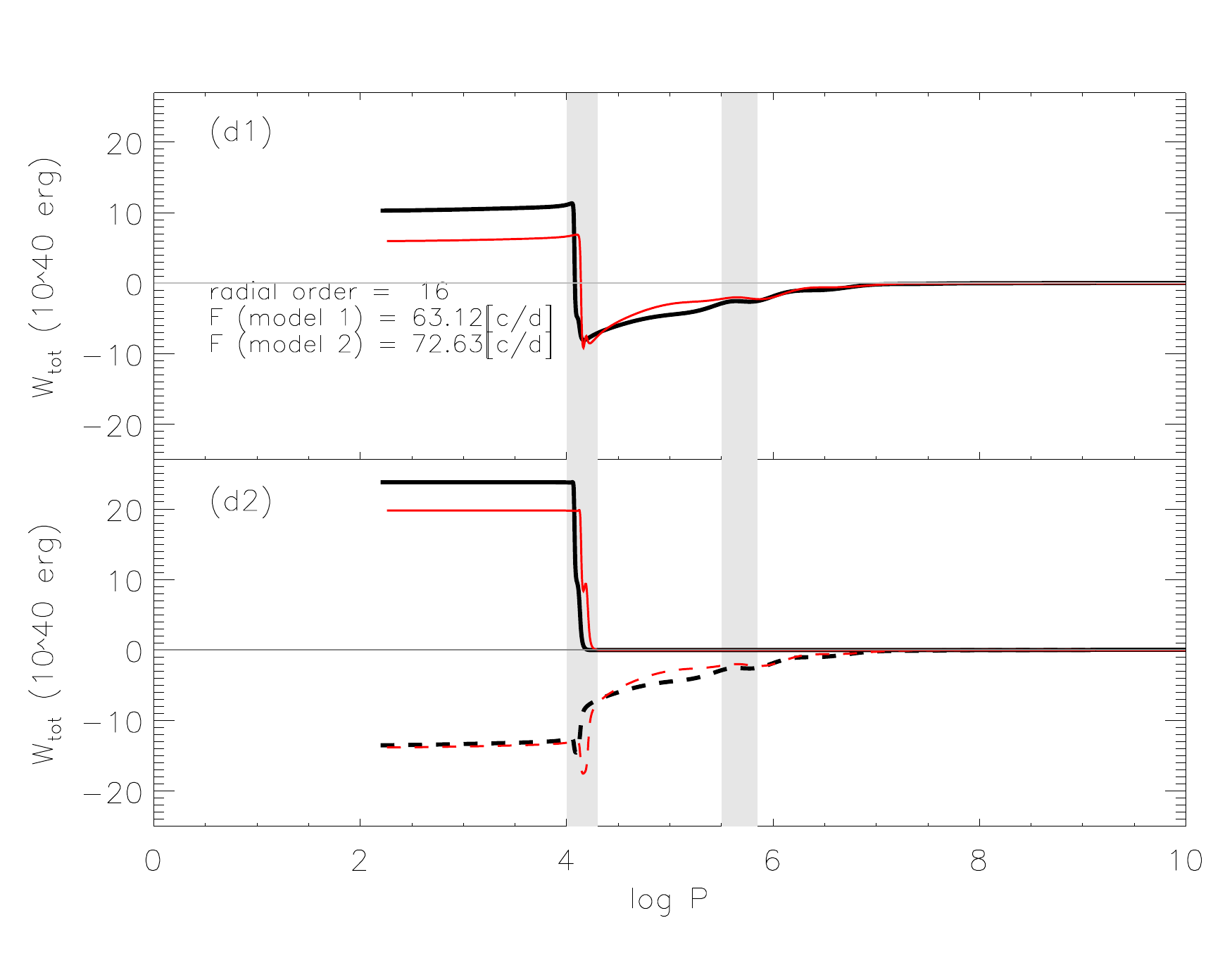}

\caption{Accumulated work integrals as a function of total pressure for modes of different radial orders $n$. Model 1 is depicted in black and model 2 in red (gray in the printed version). The parameters are described in Table 1. The grey region at log {\it P} $\backsimeq6$ indicates the He\,II ionization region, and at log {\it P} $\backsimeq4$ the H\,I and He\,I ionization zones. Panels (a1), (b1), (c1) and (d1) illustrate the total accumulated work integrals, while panels  (a2), (b2), (c2) and (d2) show the accumulated work integrals for the turbulent pressure (continuous line) and the gas pressure (dashed line).   \label{work}}

\end{center}

\end{figure*}

Inspection of the work integrals\footnote{ A detailed discussion of the work integrals $W_{\rm t}$ and $W_{\rm g}$ is given in, e.g., Balmforth 1992a.} for these modes shows that it is the
perturbation\footnote{We use the word $perturbation$
to denote a variation of a quantity induced by the pulsation and the word $fluctuation$ 
for the variation of a quantity induced by the convective motion.} of the turbulent pressure (or momentum flux) which is responsible
for driving most of the high-frequency modes. This is reflected in a local increase of the accumulated work integral contribution $W_{\rm t}$
(solid curve) in Fig.~\ref{work}. The contribution from the gas pressure to the
accumulated work integral, $W_{\rm g}$ (dashed curves in Fig.~\ref{work}), which includes
contributions from both the radiative and convective heat flux, also increases locally 
in the H\,I ionization zone, but by a much smaller amount than $W_{\rm t}$.
In Fig.~\ref{work} we show the accumulated work integrals for four different radial orders, $n$=4, 7, 12 and 16 for both models explored in this article. The mode with n=4 corresponds roughly to the dominant mode observed in HD\,187547, which is at approximately 21\,cd$^{-1}$. The positive accumulated work integral in Fig.~\ref{work},  panel (a1) shows that this mode is primarily excited by gas pressure in the He\,II ionization zone (see panel (a2) in Fig.~\ref{work}), i.e. by the `classical' opacity mechanism. In this case the turbulent pressure has a slight damping effect. For the mode with $n$=7, however, driving in the hydrogen zone occurs by both the gas and the turbulent pressure, but in the He\,II ionization region only the opacity mechanism is effectively contributing to driving (Fig.~\ref{work}, panels (b1), (b2)). Interestingly, for the $n$=12 and 16 modes, depicted in Fig.~\ref{work} panels (c1,2) and (d1,2) respectively, the gas pressure is entirely damping while the turbulent pressure in the H\,I/He\,I ionization zones is responsible for driving these modes.  

 The turbulent pressure in the He\,II ionisation layer does not play an important role neither for the low- nor the high- order modes because, additionally to the very inefficient convection in this ionisation zone, the pulsation periods are much shorter compared to the characteristic convection time-scale $\tau$. In the H ionisation zone, however, $\tau$ is similar to the pulsation periods for frequencies between 60 and 80\,cd$^{-1}$ (650 and 800\,$\mu$Hz), which is also reflected in Fig.~\ref{growth} by the largest values of the linear growth rates (black and red solid curves).

Similar to the case of the $\kappa$ mechanism, which 
is responsible for driving the lowest radial orders in the deeper He\,II 
ionization region, the driving by the convection dynamics occurs through the pulsationally induced turbulent pressure perturbations
$\delta p_{\rm t}$, where
$p_{\rm t}:=\langle\rho ww\rangle$ with $\rho$ being density
and $w$ the vertical component of the convective velocity
field ${\bm u} = ({\it u, v, w})$ (angular brackets denote
an ensemble average, but in practice horizontal
averages are used) and $\delta$ denotes a perturbation
in a Lagrangian frame of reference. This leads to intrinsically unstable (self-excited) modes, i.e. to 
{\it coherent} modes, which are consistent with the narrow widths of the
high-frequency peaks in the observed acoustic oscillation spectrum 
derived from the longer Kepler data set presented here.

There are two distinct ways for the turbulent pressure $p_{\rm t}$ to excite waves. Solar-like, or stochastic, excitation is caused by localised acoustic random events in the fluctuations of the Reynolds stresses $\rho\vec{u}\vec{u}$. This excitation is constrained to a rather thin layer, just below the stellar surface, where convection reaches the highest turbulent Mach numbers. These convective fluctuations have random phases with respect to the modes and therefore constitute a stochastic driving. A rather different mechanism is offered by the coherent excitation by the horizontally averaged turbulent pressure $p_{\rm t}$ due to a phase-lag in the response of $p_{\rm t}$ to an incident (acoustic) wave or density perturbation, similar to the phase-lag
between gas pressure $p$ and density $\rho $ in the case of the $\kappa$ mechanism. The phase lag is such that it provides either coherent excitation, or damping, but does not fluctuate in time, as is the case in the stochastic excitation mechanism. Instead it depends only on the structure of the star.

\begin{table}[b!]
\caption{ Parameters for the models used in this work are consistent with the observations and can explain the major part of the observed modes. \label{data}}
\begin{center}
\begin{tabular}{ccccc} \hline
  & Mass &$T_{\rm eff}$ &L &   $\alpha$\\
  &[$M/M\sun$]& [K]          &  [$L/L\sun$] & \\\hline
 model 1 & 1.85 & 7575 & 16 &1.7 \\ 
 model 2 & 1.85 & 7575 & 19 & 1.7\\ \hline%\\%[3pt]
  %              &                                &  &&&
%\hline

 \end{tabular}

 \end{center}
\end{table}

Previous calculations with a time-dependent convection formulation already 
indicated that the turbulent pressure can act as a driving agent in various
types of pulsating stars. The first evidence that the phase of $\delta p_{\rm t}$ can be such as to drive pulsations
was reported by Gough (1966) for Mira stars.
Later Houdek (2000) found intrinsically unstable 
modes with radial orders $7\la n\la 10$ in $\delta$ Scuti stars for a certain
set of convection parameters and, more recently, Xiong et al. (2007) reported an instability strip for red giants whose location in the 
HR-diagram was determined by the turbulent pressure contribution 
to the work integral.

We conclude that driving by turbulent pressure 
perturbations in an appropriate stellar model does reproduce up to 85\% of the the observed 
range of unstable modes in HD\,187547. This driving mechanism is therefore a
promising candidate for explaining high-order p modes in $\delta$ Scuti stars. 
    
\subsection{Insights from 3D simulations of convection}

In 1D models of convection, including mixing length treatment (MLT) as well as nonlocal and
time-dependent flavours of it, the convective flux is inseparably linked to
the velocities through simple proportionality. Small convective fluxes therefore mean small convective velocities. In Gough's (1977 a, b) non-local mixing-lengths formulation, the differential equations (of 6th order) for the turbulent fluxes are solved as a boundary-value problem, with exponentially decaying (positive) fluxes into the overshooting, convectively stable layers. From realistic, 3D hydrodynamic simulations of convection (Trampedach 2013), on the other hand, 
we know that the velocity field is almost decoupled from the fluxes, and the
latter are governed not only by correlations between flow direction and
temperature contrast, but also by the fractional area (filling factor) covered by
each flow direction. Since large fluxes are going in each direction, even a slight
change in filling factor can profoundly change the convective fluxes. This is
what decouples the convective velocities from the convective fluxes, and makes
current 1D formulations of convection inadequate for predicting velocity
fields from overshooting and from very inefficient convection, as found in A
stars.

At the borders of convection zones, the Schwarz\-schild criterion tells us that
the correlation between temperature contrast and flow direction
changes sign, and the overshoot flux is actually negative, as
observed in the simulations. This overshoot flux is also small as the upflow
filling factor abruptly changes from 65\% in the convection zone to less than 50\% 
above it, causing the opposing fluxes to nearly cancel and fall off
exponentially with height. The velocities,
on the other hand, display no sign of this profound transition. Since the turbulent pressure is
$\langle\varrho ww\rangle$ and $\varrho$ does not change abruptly either,
the turbulent pressure extends well beyond the convection zones,
despite the fluxes forming a well-defined boundary of convection.
Thin and close convection zones, as found in A stars, therefore have
efficient mixing between them, despite the two zones being clearly
separated. They also have very inefficient convection, with high
velocities transporting only a fraction of the full flux, $\sigma T_{\rm eff}^4$ (Freytag, Ludwig, \& Steffen 1996, Freytag \& Steffen 2004, Trampedach 2004, Kupka, Ballot \& Muthsam 2009).

The decline of convective fluxes occurs at cooler temperatures compared to 
the decline of the convective velocities, whereas in the case of 1D MLT models, the velocities and fluxes are proportional to each other. In other words, the outer layers of A-type stars are still convective. In the case of high effective temperatures, however, convection mixes material but does not transport energy efficiently. Despite the free parameters of analytical formulations of
convection, no combinations of parameters can reproduce these results of
3D simulations, due to their built-in assumptions. Compared to such a model
a preliminary convection simulation of HD\,187547 has a larger and more
extended turbulent pressure peak, providing even more excitation than the 1D
model, despite transporting little flux by convection.

\section{Discussion \& Conclusion}

Based on more than two years of {\it Kepler} data we can conclude that  the high-radial order modes observed in HD\,187547, initially interpreted by Antoci et al. (2011) as being stochastically excited, show no temporal variability and are coherent and intrinsically unstable (i.e. self-excited). The long data set was required to resolve the very closely spaced peaks and to show that there is no temporal variability, {\it inconsistent with pure stochastic oscillations that have mode lifetimes significantly shorter than the observing period of 960 days}. \par
There is no clear explanation for the observed agglomeration of peaks around the high radial order modes, as we cannot account for all the peaks using only modes with $l\le3$ and their associated $m$ modes. Modes with $l=4$ are not expected to cluster around high-order modes with $l\le3$, at least not for the observed parameters of HD\,187547 and are thus also insufficient. We cannot exclude the presence of modes with $l>4$, but the geometrical cancelation effects for such modes are very high, which would imply that the intrinsic amplitude of these high-degree modes are higher than those of radial or dipolar modes -- a hypothesis that we find unlikely and that has no other observational support. From polarimetric observations we set an upper limit of 20 G for the strength of the magnetic field, which excludes the possibility of magnetically split modes like those seen in roAp stars. \par

Our models using time-dependent, nonlocal convection treatment suggest that convection excites oscillations in HD\,187547, however, not stochastically. It is the turbulent pressure, that is the main driving agent, and can reproduce 85\% of the observed frequency range. We find that the adopted nonlocal convection parameters, $a^2=950$, $b^2=950$ and $c^2=950$ can reproduce observations of oscillating A-type stars and therefore assume that they represent realistic values for describing the convection dynamics in A type stars when using the nonlocal convection model by Gough (1977). Preliminary results from 3D hydrodynamic simulations of convection for HD\,187547 show that the amount of turbulent pressure is even larger than in our 1D models, which provides further evidence in favour of this excitation mechanism provided the turbulent pressure has a suitable phase lag with respect to the oscillations.

Similar values for the $a$, $b$ and $c$ parameters, as adopted in this work, were used in the study of the excitation mechanism in roAp stars conducted by Balmforth et al. (2001), equally motivated by the fact that in these pulsators the convective envelope is very shallow. Balmforth et al (2001) found that driving of high-frequency modes in roAp stars takes place in the H\,I ionization region, however, the driving results from the opacity mechanism and is present only in models where convection is suppressed by a magnetic field. Thus, the values of $a$, $b$, and $c$ play no direct role in the excitation of the oscillations discussed by Balmforth et al (2001). Nevertheless, based on a detailed comparison with observations, Cunha et al. (2013) recently provided strong evidence that the opacity mechanism in models similar to those suggested by Balmforth et al. (2001) fails to drive the very high-frequency oscillations observed in a number of roAp stars. As an alternative, Cunha et al. (2013) suggested that the perturbation to the turbulent pressure may be the principle driving agent for these modes, similarly to our independent findings for models of HD\,187547. However, since the oscillations observed in roAp stars are of significantly higher radial orders and, consequently, more sensitive to the details of the outer stellar layers  - which are particularly difficult to model in roAp stars, due to the presence of strong magnetic fields and strong chemical peculiarities - the authors were very conservative in their conclusions about this possibility.  Our results, for intermediate radial order modes in a star with a much less complex outer envelope and atmosphere, support  the possibility that the turbulent pressure can be a driving agent for classical pulsators in a particular region of the HR diagram. 

For the future we plan to explore the entire parameter space in the classical instability strip and investigate how the values of $a$, $b$, $c$ affect the mode stability within the entire instability strip, i.e. for different stellar masses, metallicities and changing the stratification in the stellar atmospheres using the models presented in this paper. Additionally we have started 3D simulations of convective atmospheric models for the parameters of HD\,187547, which will allow to better understand the stellar convective envelope in $\delta$\,Sct and related stars. 

\section{Acknowledgments}

Funding for the Stellar Astrophysics Centre is provided by The Danish National Research Foundation (Grant agreement no.: DNRF106). The research is supported by the ASTERISK project (ASTERoseismic Investigations with SONG and Kepler) funded by the European Research Council (Grant agreement no.: 267864). MSC is supported by an Investigador FCT contract funded by FCT/MCTES (Portugal) and POPH/FSE (EC) and by funds from the ERC, under FP7/EC, through the project FP7-SPACE-2012-31284. GHa acknowledges support by the Polish NCN grant 2011/01/B/ST9/05448. TL would like to thank TBL and the staff at the Pic du Midi observatory for their excellent support, in particular the availability of DDT time for carrying out the spectropolarimtric observations. TL acknowledges the support by the FWF NFN project S116601-N16 and the related FWF NFN subproject, S116 604-N16.

\clearpage

\end{document}